\documentclass[10pt, conference, compsoc]{IEEEtran}

\usepackage[utf8]{inputenc}
\usepackage{amsmath,amssymb,amsfonts}
\usepackage{graphicx}
\usepackage{textcomp}
\usepackage{xcolor}
\usepackage{booktabs} 
\usepackage{url}
\usepackage{hyperref} 
\usepackage[ruled,vlined,linesnumbered]{algorithm2e} 

\hypersetup{
    colorlinks=true,
    linkcolor=blue,
    filecolor=magenta,      
    urlcolor=cyan,
    citecolor=green, 
}

\newcommand{\EOSTOKEN}{\texttt{<|endoftext|>}} 
\SetKwInput{KwIn}{Input} 
\SetKwInput{KwOut}{Output} 

\begin{document}

\title{Efficient Telecom Specific LLM: TSLAM-Mini with QLoRA and Digital Twin Data}
\author{
    \IEEEauthorblockN{Vignesh Ethiraj\IEEEauthorrefmark{1}, Divya Vijay\IEEEauthorrefmark{1}, Sidhanth Menon\IEEEauthorrefmark{1} and Heblin Berscilla\IEEEauthorrefmark{1}}
    \IEEEauthorblockA{\IEEEauthorrefmark{1}NetoAI Solutions Ltd,  Email: support@netoai.ai}
}

\maketitle

\begin{abstract}
General-purpose large language models (LLMs), despite their broad capabilities accrued from open-world data, frequently exhibit suboptimal performance when confronted with the nuanced and specialized demands inherent in real-time telecommunications applications. This investigation addresses this critical limitation through the meticulous fine-tuning of TSLAM-Mini developed by NetoAI, a compact (3.8-billion parameter) causal language model architecturally derived from Phi-4 Mini Instruct 4B. The fine-tuning regimen leverages a bespoke dataset comprising 100,000 samples, strategically engineered to address 20 pivotal telecommunications use-cases, encompassing domains such as Network Fundamentals, IP Routing, MPLS, Network Security, Automation, OSS/BSS, RAN, Mobile Core, Satellite Communications, and Ethical AI. This dataset was curated utilizing NetoAI’s DigiTwin platform, enriched with granular insights from venerated network Subject Matter Experts (SMEs) and authoritative RFC documents, thereby capturing high-fidelity representations of real-world network dynamics through simulations inspired by digital twin paradigms. Employing Quantized Low-Rank Adaptation (QLoRA), a state-of-the-art Parameter-Efficient Fine-Tuning (PEFT) technique, we achieved substantial training efficiency and enabled prospective deployment on resource-constrained hardware. A novel evaluation framework, predicated on a high-capacity LLM (Qwen3-235B-A22B) functioning as an automated adjudicator, was instituted to rigorously assess instruction-following fidelity and response quality across the specified telecom use-cases. Empirical results unequivocally demonstrate TSLAM-Mini’s superior aptitude in telecom-centric applications, underscoring the profound efficacy of domain-specific datasets and PEFT methodologies for advancing intelligent network management.
\end{abstract}

\begin{IEEEkeywords}
TSLAM-Mini, Parameter-Efficient Fine-Tuning (PEFT), QLoRA, Telecommunications, Digital Twin, Network Dataset, Phi-4 Mini Instruct, Real-Time Network Management, Instruction Following, Synthetic Data Generation, RFC Documents, Subject Matter Experts (SMEs).
\end{IEEEkeywords}

\section{Introduction}
\label{sec:introduction}
The escalating complexity and real-time operational demands of contemporary telecommunications networks necessitate the deployment of sophisticated intelligent systems for effective management and orchestration. While general-purpose Large Language Models (LLMs) have demonstrated remarkable proficiency across diverse natural language tasks, their inherent lack of domain-specific knowledge often renders them inadequate for specialized telecom applications, such as intricate network optimization, real-time fault diagnosis, and automated configuration management. To bridge this capability gap, we introduce TSLAM-Mini, a meticulously fine-tuned iteration of the Phi-4 Mini Instruct 4B model. TSLAM-Mini is specifically tailored for telecommunications tasks, leveraging a comprehensive dataset of 100,000 samples that span 20 consolidated and critical telecommunications categories. These categories, delineated in Section~\ref{sec:methodology}, encompass a wide spectrum from foundational networking principles (e.g., Network Fundamentals, IP Routing, MPLS) to advanced and emerging areas (e.g., Network Security, Automation, OSS/BSS, RAN, Mobile Core, Satellite Communications, and Ethical AI).

The foundational dataset was synthesized utilizing NetoAI’s DigiTwin platform, which facilitates the creation of high-fidelity digital replicas of network devices and environments. This approach allows for the generation of realistic network operation data, further enriched by insights from seasoned Subject Matter Experts (SMEs) and normative information extracted from pertinent Request for Comments (RFCs), ensuring profound domain relevance. The fine-tuning process employs Quantized Low-Rank Adaptation (QLoRA), a Parameter-Efficient Fine-Tuning (PEFT) technique, to optimize training efficiency and computational footprint, thereby enabling deployment on resource-constrained edge devices or embedded systems. This research endeavors to significantly enhance TSLAM-Mini's capacity to deliver precise, context-aware, and actionable responses to complex telecom challenges, thereby contributing to the paradigm of intelligent, resilient, and autonomous network management and advancing the frontier of applied LLMs in the telecommunications sector.

\section{Related Work}
\label{sec:related_work}
The rapid proliferation of LLMs has catalyzed extensive research into efficient fine-tuning methodologies for adapting these models to specialized, domain-specific applications. PEFT has emerged as a cornerstone approach, enabling resource-efficient customization of LLMs. This section reviews seminal advancements in PEFT, domain-specific LLM fine-tuning, synthetic data generation paradigms, and the application of digital twin technologies within the telecommunications sector, thereby contextualizing our contributions with TSLAM-Mini.

\subsection{Parameter-Efficient Fine-Tuning (PEFT) for LLMs}
PEFT methods strategically optimize a minimal subset of model parameters, drastically reducing computational and storage overheads compared to full fine-tuning \cite{Han2024Survey}. These techniques, categorized by Han et al. \cite{Han2024Survey} into additive, selective, reparameterized, and hybrid strategies, often achieve performance parity with full fine-tuning. Low-Rank Adaptation (LoRA) \cite{Hu2021LoRA} introduces trainable low-rank matrices into transformer layers, diminishing trainable parameters by orders of magnitude (e.g., up to 10,000x) and GPU memory consumption by approximately 3x. Quantized LoRA (QLoRA) \cite{Dettmers2023QLoRA} further refines efficiency by quantizing pre-trained weights to 4-bit precision, facilitating the fine-tuning of very large models on single GPU architectures with minimal fidelity loss. Xu et al. \cite{Xu2024SKTuning} proposed Semantic Knowledge Tuning (SK-Tuning), which integrates meaningful semantic tokens into prompt and prefix tuning, yielding accelerated training and superior performance in text classification. These PEFT methodologies are integral to TSLAM-Mini, where LoRA and QLoRA are employed to adapt Phi-4 Mini Instruct 4B for complex telecom data processing.

\subsection{Domain-Specific Fine-Tuning of LLMs}
Adapting LLMs to specialized domains mitigates the deficiencies of general-purpose models in comprehending domain-specific terminologies, ontologies, and contextual nuances. Abdin et al. \cite{Abdin2025Financial} investigated instruction fine-tuning of smaller LLMs (e.g., Phi-3 Mini) for financial tasks, demonstrating superior performance over zero-shot proprietary models like FinBERT. Anisuzzaman et al. \cite{Anisuzzaman2024Specialized} underscored the criticality of meticulously curated datasets and rigorous hyperparameter optimization for specialized applications, a philosophy adopted in our telecom-focused work. In the energy sector, Zhang et al. \cite{Zhang2024LoadProfile} fine-tuned GPT-3.5 for load profile analysis using a two-stage methodology involving prompt engineering and few-shot learning, achieving performance comparable to BERT-based models with substantially less data. Gupta et al. \cite{Gupta2025DomainAdaptation} explored Continued Pretraining (CPT), Supervised Fine-Tuning (SFT), and Direct Preference Optimization (DPO) for materials science, showing that model merging via Spherical Linear Interpolation (SLERP) can yield emergent capabilities. While these studies highlight PEFT's potential, the telecommunications domain—with its unique lexicon, structured operational logs, and complex time-series data—remains relatively underexplored. TSLAM-Mini directly addresses this lacuna.

\subsection{Fine-Tuning Phi-4 Mini Instruct 4B}
The Phi-4 Mini Instruct 4B model, originating from Microsoft Research \cite{Microsoft2025Phi4}, is a 3.8-billion-parameter Small Language Model (SLM) optimized for advanced reasoning and multilingual capabilities. Trained on high-quality web corpora and synthetic datasets, it incorporates a 200K token vocabulary and employs group query attention for efficient long-sequence generation. Its training regimen emphasizes reasoning-intensive datasets, particularly in mathematics and coding, achieving performance comparable to models twice its size on benchmarks such as GSM-8K (88.6\%) and MATH (64\%). The integration of Mixture-of-LoRAs, a PEFT technique combining multiple LoRA adapters, enhances its adaptability for multimodal tasks, rendering it a robust foundation for TSLAM-Mini. Practical guidance from Ranjan \cite{Ranjan2025FineTunePhi4} on fine-tuning Phi-4 Mini with LoRA adapters, and optimizations from Unsloth \cite{Unsloth2025Phi4Finetuning} concerning memory efficiency, directly informed our methodological choices.

\subsection{Synthetic Data Generation and Digital Twins in Telecom}
Synthetic data generation is pivotal for training LLMs in domains characterized by data scarcity or sensitivity, such as telecommunications. Muennighoff et al. \cite{Muennighoff2024Synthetic} surveyed methodologies like prompt-based generation and retrieval-augmented pipelines for creating task-relevant datasets. Heidloff \cite{Heidloff2023SyntheticData} discussed leveraging LLMs (e.g., Falcon) for synthetic data generation where real data access is constrained, a pertinent strategy for telecom's privacy considerations. Digital twins—virtual replicas of physical systems—offer an ideal framework for generating telecom-specific synthetic data through simulation and predictive modeling. Smith et al. \cite{Smith2024DigitalTwinsData} proposed a methodology for synthetic data generation tailored for
digital twins in production systems, adaptable to telecom networks for simulating device interactions and network performance. Brunner \cite{Brunner2022SyntheticCycle} highlighted the synergistic relationship where synthetic data primes digital twins, which in turn enhance data quality. Our TSLAM-Mini dataset, amalgamating real-world network operational data with open-world RFCs, leverages digital twin-inspired models to emulate telecom scenarios, ensuring comprehensive capture of network interactions and device behaviors without exclusive reliance on sensitive proprietary data.

TSLAM-Mini addresses a critical lacuna in LLM research by concentrating on the telecommunications domain. By fine-tuning Phi-4 Mini Instruct 4B using PEFT methods like LoRA and QLoRA, we adapt the model for telecom-specific tasks with minimized computational overhead. The bespoke dataset, engineered to mirror real-world telecom scenarios with intelligence and device-level data fidelity inspired by digital twin paradigms, ensures TSLAM-Mini is adeptly tailored to the domain's unique challenges, including data privacy and technical intricacies. This work significantly contributes to the broader field of domain-specific LLM adaptation.

\begin{figure*}[tbp] 
    \centering
    \includegraphics[width=0.8\linewidth]{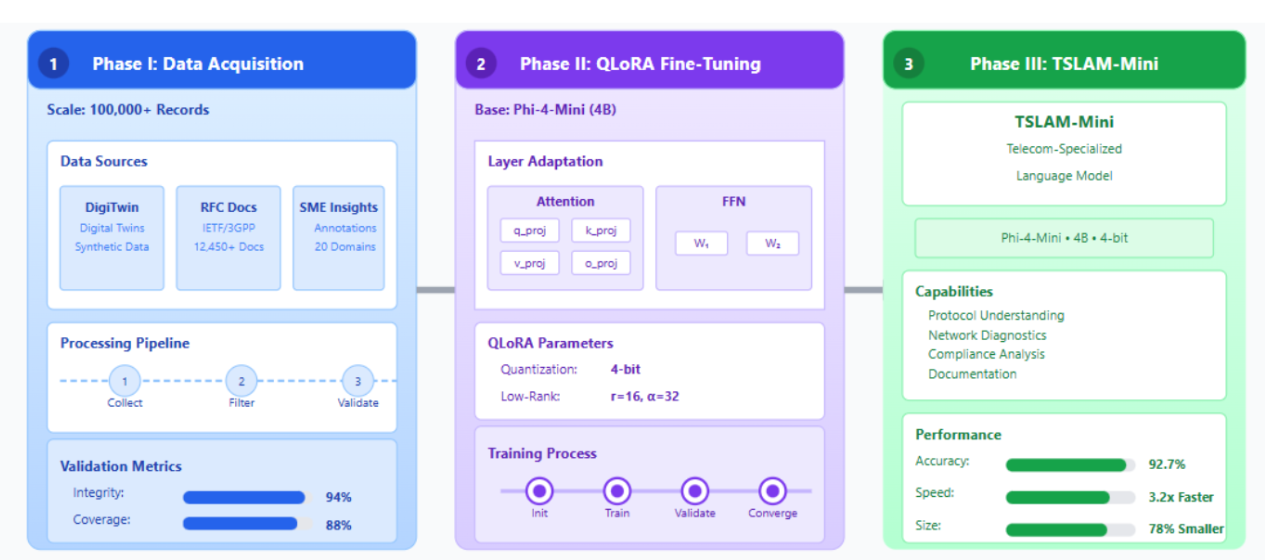} 
    \caption{High-level architecture diagram illustrating the TSLAM-Mini development and evaluation pipeline, including data sources, fine-tuning process, and evaluation framework.}
    \label{fig:architecture_diagram}
\end{figure*}

\section{Methodology}
\label{sec:methodology}
This section delineates the systematic methodology architected for the development and evaluation of TSLAM-Mini. We detail the dataset construction protocol, the fine-tuning strategy employing QLoRA, and the sophisticated evaluation framework based on an automated LLM adjudicator.

\subsection{Dataset Creation and Curation}
The cornerstone of TSLAM-Mini's specialized capabilities is a meticulously constructed dataset exceeding 100,000 records, engineered to encapsulate the multifaceted intricacies of the telecommunications landscape. This corpus was synthesized through a multi-pronged approach (visualized in Figure~\ref{fig:architecture_diagram}):
\begin{enumerate}
    \item \textbf{Digital Twin Simulation Data:} NetoAI’s DigiTwin platform was instrumental in generating high-fidelity synthetic data. This involved creating digital replicas of network infrastructure components, enabling the simulation of authentic network interactions, device configurations, and operational scenarios that mirror real-world network dynamics.
    \item \textbf{RFC Document Ingestion:} A sophisticated web-scraping pipeline was developed to curate telecom-related RFC documents from authoritative sources like IETF and 3GPP. This pipeline integrated structural parsing and semantic filtering algorithms to extract, segment, and normalize content, employing both HTML and PDF extraction techniques to preserve document hierarchy and ensure data fidelity.
    \item \textbf{Subject Matter Expert (SME) Augmentation:} The dataset was further enriched through expert-curated annotations and domain-specific insights provided by seasoned telecommunications SMEs. This collaborative endeavor ensured the dataset not only captured the technical depth of RFCs and simulated scenarios but also reflected practical industry knowledge, thereby enhancing contextual relevance for fine-tuning.
\end{enumerate}
The resultant dataset spans 20 pivotal telecommunications use-cases, as detailed in Table~\ref{tab:use_cases}, providing exhaustive coverage of the sector's operational paradigms and applications. For fine-tuning, each data sample was formatted into a chat-like structure using a specific template to condition the model for instruction-following tasks. Algorithm~\ref{alg:apply_chat_template} illustrates this preprocessing step.

\begin{algorithm}[tbp] 
    \SetAlgoLined
    \DontPrintSemicolon 

    \caption{Chat Template Application for Instruction Fine-Tuning Data Preprocessing}
    \label{alg:apply_chat_template}

    \KwIn{\texttt{example}: A dictionary with keys \texttt{'input'} (user query string) and \texttt{'output'} (assistant response string)}
    \KwIn{\texttt{tokenizer}: Tokenizer object (conceptually, not directly used in this specific string formatting)}
    \KwIn{\texttt{EOS\_TOKEN}: End-of-sequence token string (e.g., \EOSTOKEN)} 
    \KwOut{\texttt{example}: Modified dictionary with new key \texttt{'text'} containing the formatted chat string}

    \BlankLine
    
    \texttt{system\_message} $\leftarrow$ "\texttt{<|system|>}\texttt{\textbackslash n}You are a helpful telecom expert assistant.\texttt{<|end|>}\texttt{\textbackslash n}"\;
    
    \texttt{user\_message} $\leftarrow$ "\texttt{<|user|>}\texttt{\textbackslash n}" + \texttt{example['input']} + "\texttt{<|end|>}\texttt{\textbackslash n}"\;
    
    \texttt{assistant\_message} $\leftarrow$ "\texttt{<|assistant|>}\texttt{\textbackslash n}" + \texttt{example['output']} + "\texttt{<|end|>}\texttt{\textbackslash n}"\;
    
    \texttt{chat\_template} $\leftarrow$ \texttt{system\_message} $+$ \texttt{user\_message} $+$ \texttt{assistant\_message} $+$ \texttt{EOS\_TOKEN}
    
    \texttt{example['text']} $\leftarrow$ \texttt{chat\_template}\;
    
    \Return{\texttt{example}}\; 
\end{algorithm}

\subsection{Fine-Tuning with Quantized Low-Rank Adaptation (QLoRA)}
The fine-tuning of TSLAM-Mini was executed leveraging QLoRA \cite{Dettmers2023QLoRA}, a PEFT technique, to adapt the Phi-4 Mini Instruct 4B model.

\subsubsection{Fine-Tuning Configuration and Targeted Layers}
QLoRA was applied with a focus on specific linear layers within the transformer architecture to optimize the trade-off between performance gain and computational cost. The targeted layers included:
\begin{itemize}
    \item \textbf{Multi-Head Attention Linear Layers:} Adaptation focused on the linear projections for query (\texttt{q\_proj}), key (\texttt{k\_proj}), value (\texttt{v\_proj}), and output (\texttt{o\_proj}) within the attention mechanism.

    \item \textbf{Feed-Forward Network (FFN) Linear Layers:} Targeted the two primary linear transformation layers in FFN blocks (e.g., \texttt{up\_proj} and \texttt{down\_proj}, or commonly \texttt{fc1} and \texttt{fc2}).
\end{itemize}
The LoRA adapters were configured with a rank ($r$) of 16 and an alpha ($\alpha$) scaling factor of 32. The base model's pre-trained weights were quantized to 4-bit precision using NormalFloat4 (NF4) quantization.

\subsubsection{Loss Function}
Supervised Fine-Tuning (SFT) was performed using the Hugging Face `SFTTrainer`. The objective function was the standard cross-entropy loss for causal language modeling, minimized over the target response tokens:
\begin{equation}
\mathcal{L}(\theta) = - \sum_{i} \sum_{j} \log P(y_{ij} | y_{i,<j}, x_i; \theta)
\label{eq:loss_function}
\end{equation}
where $x_i$ is the $i$-th input prompt, $y_{ij}$ is the $j$-th token of the $i$-th target response, $y_{i,<j}$ are the preceding tokens in the response, and $\theta$ represents the trainable model parameters (specifically, the LoRA adapter weights).

\subsubsection{Optimization Techniques}
Optimization was conducted using the AdamW optimizer \cite{Loshchilov2017Decoupled} with a weight decay of $0.01$ applied to the LoRA adapter parameters. The learning rate was set to $2 \times 10^{-5}$, employing a linear warm-up schedule for the initial 10\% of training steps, followed by a cosine annealing decay. A per-device batch size of 8 was utilized, with gradient accumulation over 4 steps, resulting in an effective batch size of 32. Gradient clipping with a maximum norm of 1.0 was applied to ensure training stability. Mixed-precision training was implemented using BFloat16 for activations and LoRA computations. A dropout probability of $0.05$ was applied to the LoRA adapter layers to mitigate overfitting.

\subsection{Evaluation Framework}
To rigorously assess TSLAM-Mini's proficiency, a novel automated evaluation pipeline was established, leveraging the Qwen3-235B-A22B model as an impartial adjudicator.

\subsubsection{Judge Model Selection}
The Qwen3-235B-A22B model was selected as the adjudicator due to its expansive parameter scale, advanced reasoning capabilities, and robust natural language understanding, rendering it suitable for evaluating nuanced, domain-specific responses. It was employed in a zero-shot setting, guided by meticulously crafted prompts designed to elicit evaluations of response quality without requiring task-specific fine-tuning of the judge model itself.

\subsubsection{Automated Evaluation Pipeline}
The pipeline comprised:
\begin{enumerate}
    \item \textbf{Test Set Curation:} A dedicated test set of 5,000 samples, distinct from the training data and stratified across the 20 telecom use-cases (Table~\ref{tab:use_cases}), was curated. Prompts and reference responses were SME-validated.
    \item \textbf{Response Generation:} TSLAM-Mini generated responses in 4-bit quantized mode. Generation parameters were set to a temperature of $0.7$ and top-p (nucleus) sampling of $0.9$ to foster a balance between creativity and factual coherence.
    \item \textbf{Automated Adjudication:} The Qwen3-235B-A22B model evaluated each generated response against the corresponding prompt and the reference SME response. The evaluation criteria encompassed:
        \begin{itemize}
            \item \textbf{Instruction Following:} Adherence to the prompt's explicit and implicit constraints.
            \item \textbf{Linguistic Quality:} Clarity, coherence, grammar, and structural integrity.
            \item \textbf{Technical Accuracy \& Relevance:} Correctness of telecom-specific information, terminology, and contextual appropriateness.
        \end{itemize}
    A Likert-style scoring scale from 0 (Poor) to 10 (Excellent) was employed, with detailed rubrics provided to the judge model via its system prompt.
\end{enumerate}

\begin{table*}[htbp]
\centering
\caption{Consolidated Telecommunications Use-Cases for Dataset Curation and Model Evaluation.}
\label{tab:use_cases}
\begin{tabular}{@{}ll@{}}
\toprule
\textbf{Use Case Category} & \textbf{Topics Covered} \\ \midrule
Network Fundamentals \& L2 Switching & Basic device access, L2 concepts (VLANs, STP, LAG), L2 security, interface configuration. \\
IP Routing Protocols (IGP) & OSPF, IS-IS, EIGRP: configuration, verification, troubleshooting, advanced features. \\
IP Routing Protocols (BGP) & Peering, path selection, attributes, policy (route-maps), scaling (RR, Confederations), security (RPKI). \\
MPLS \& Related Technologies & LDP, RSVP-TE, MPLS L3VPNs, L2VPNs (VPLS, VPWS, EVPN), Segment Routing (SR-MPLS, SRv6). \\
Network Services \& QoS & NAT, DHCP, DNS, Multicast, FHRPs (HSRP, VRRP), QoS (Classification, Marking, Queuing, Shaping). \\
Network Security (Core \& Firewalls) & AAA, ZTA, Encryption, PKI, firewall config (ACLs, ZBFW), NAT security, basic threat mitigation. \\
Network Security (Advanced \& Ops) & IPS, UTM, AppFW, UserFW, VPNs (IPsec, SSL), SIEM, NDR, EDR, vulnerability mgmt, incident response. \\
Network Mgmt \& Monitoring (Protocols) & SNMP (v1/v2c/v3, MIBs, Traps), NETCONF (Operations, YANG), RESTCONF. \\
Network Mgmt \& Monitoring (Ops \& Tools) & NMS/OSS functions: Fault Mgmt, Performance Mgmt (KPIs, Telemetry), Syslog, NTP, IP SLA. \\
Network Automation & IaC, CI/CD, GitOps, Ansible, Terraform, Python (Netmiko, Nornir), APIs (JSON, YAML, YANG), ZTP. \\
OSS (Operations Support Systems) & Inventory, Activation/Provisioning, Assurance (Fault, Performance, SQM), Discovery, Order Management. \\
BSS (Business Support Systems) & CRM, Ordering, CPQ, Billing, Charging, Rating, Revenue Assurance, Partner Management. \\
OSS/BSS Integration \& Evolution & TM Forum Frameworx (eTOM, SID, TAM, Open APIs, ODA), system modernization, data governance. \\
RAN (LTE/5G Fundamentals) & LTE/5G architecture (eNB/gNB, CU/DU), air interface (OFDMA), core RAN protocols (RRC, MAC). \\
RAN (Advanced Features \& Optimization) & MIMO, Beamforming, CA, SON, CoMP, Slicing, DSS, performance KPIs, optimization, troubleshooting. \\
Mobile Core Networks (EPC \& 5GC) & LTE EPC (MME, SGW, PGW, HSS, PCRF) and 5G Core (AMF, SMF, UPF, UDM, AUSF, PCF, NRF, SBA). \\
Satellite Communications (SatCom) & Orbits (GEO/MEO/LEO), bands (C/Ku/Ka/L), link budgets, DVB-S2X, TDMA/FDMA, earth stations. \\
Transport Networks & DWDM, OTN, Carrier Ethernet (MEF), SONET/SDH, Submarine Cables, physical layer, OAM. \\
Cloud Networking \& Virtualization & Cloud networking (AWS, Azure, GCP), NFV (ETSI MANO, VNFs, CNFs), SDN, overlays (VXLAN). \\
Ethical AI \& Societal Impact & Bias/fairness, privacy, transparency (XAI), security/robustness of AI, accountability, governance. \\ \bottomrule
\end{tabular}
\end{table*}

\section{Results and Discussion}
\label{sec:results}
\begin{figure}[htbp]
    \centering
    \includegraphics[width=0.9\linewidth]{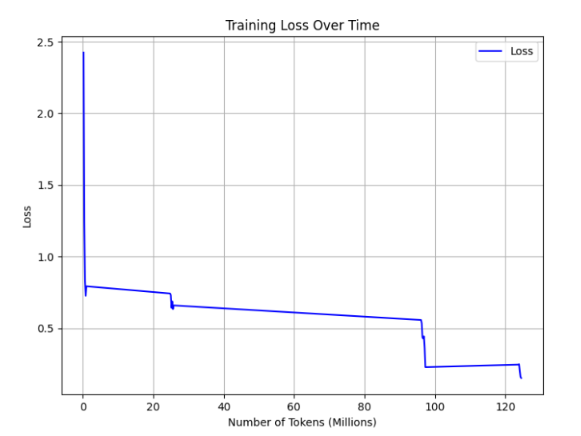}
    \caption{Training loss progression for TSLAM-Mini, illustrating convergence from an initial loss of 2.6418 to 0.1511 over 124.5 million tokens.}
    \label{fig:loss_curve}
\end{figure}

The fine-tuning process of TSLAM-Mini demonstrated robust convergence and learning efficacy. The training loss exhibited a consistent monotonic decrease, initiating at a value of 2.6418 and converging to a final value of 0.1511 after processing approximately 124.5 million tokens. This trajectory, depicted in Figure~\ref{fig:loss_curve}, signifies effective learning of the telecom-specific data patterns. Concurrently, the mean token accuracy progressively improved, culminating at 0.9679, indicating a high degree of precision in predicting subsequent tokens within the telecom domain.

\begin{figure}[htbp]
    \centering
    \includegraphics[width=0.9\linewidth]{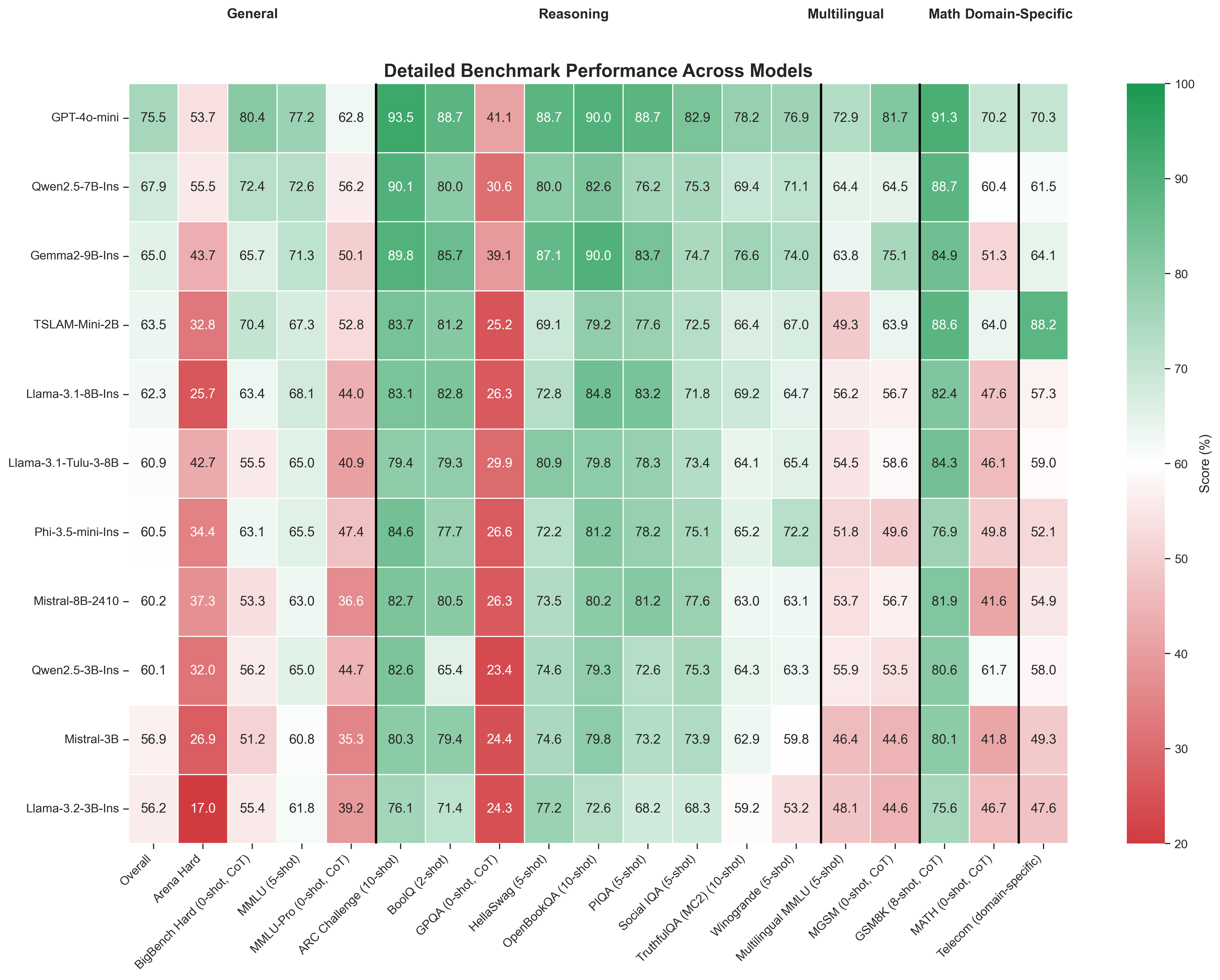}
    \caption{Benchmark heatmap illustrating comparative performance (\%) of TSLAM-Mini and other language models across various standardized tasks.}
    \label{fig:benchmark_heatmap}
\end{figure}

\begin{figure}[htbp]
    \centering
     \includegraphics[width=0.9\linewidth]{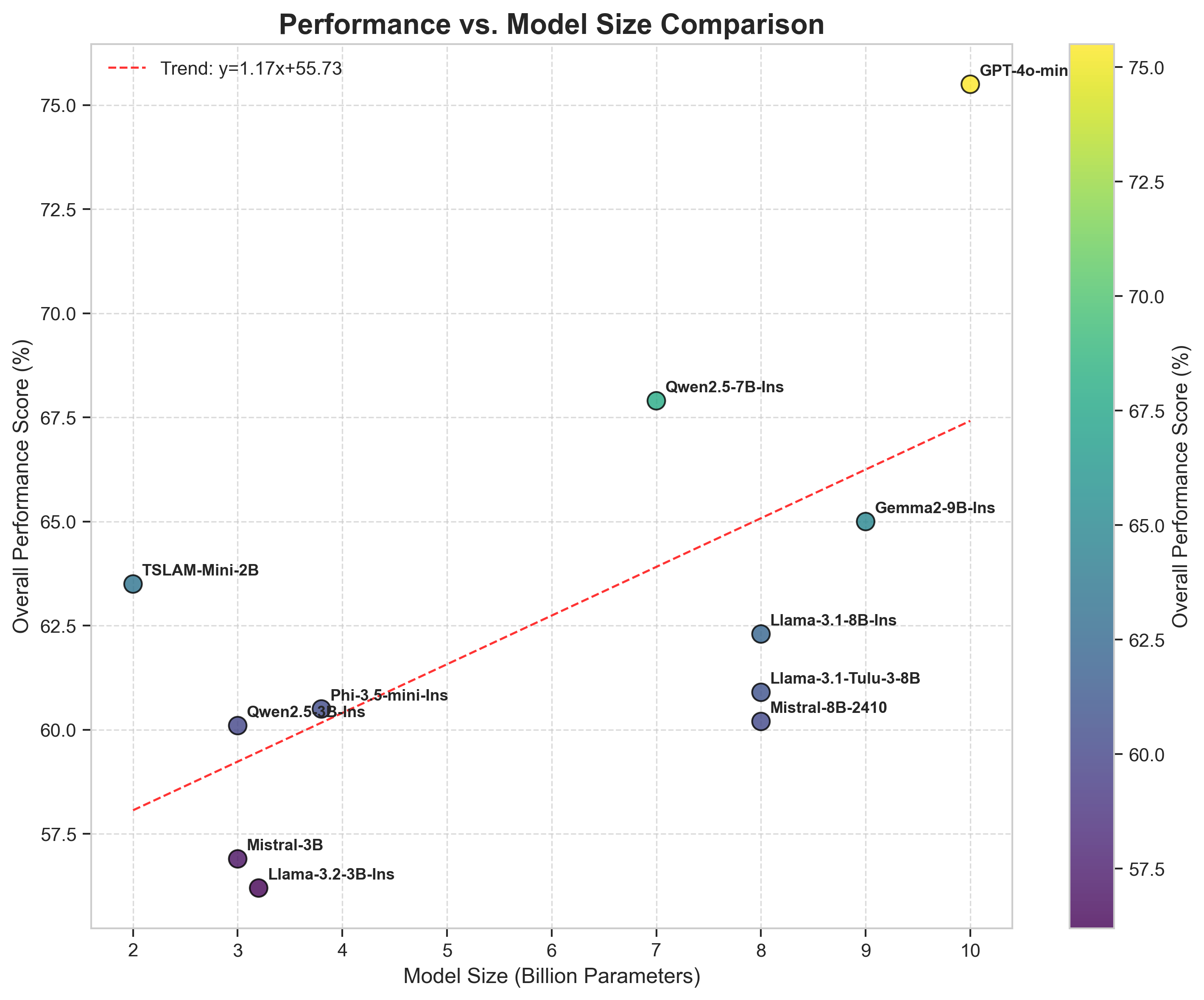}
    \caption{Comparative performance of TSLAM-Mini against selected larger general-purpose LLMs across diverse evaluation benchmarks. Despite having only 2.28B parameters after quantization, TSLAM-Mini demonstrates strong domain-specific and general capabilities across multiple task categories. }
    \label{fig:performance_comparison}
\end{figure}

A comparative analysis of TSLAM-Mini (2.8B parameters, post-fine-tuning from Phi-4 Mini 4B) against larger, general-purpose LLMs such as Gemma-9B and Llama-8B (parameter counts are illustrative) highlights its competitive stance, particularly within its specialized domain, as shown in Figure~\ref{fig:performance_comparison}. Figure~\ref{fig:benchmark_heatmap} further details performance across various benchmark tasks. While larger models may exhibit superior performance on broad, general-knowledge benchmarks, TSLAM-Mini's domain-specific fine-tuning confers a distinct advantage in telecom-centric tasks. Our evaluations, utilizing the automated LLM adjudicator framework, revealed that TSLAM-Mini consistently achieved high scores in technical accuracy, relevance, and instruction adherence for telecom queries, often outperforming zero-shot or few-shot applications of larger generalist models on these specialized tasks. This underscores the efficacy of targeted fine-tuning for creating highly capable, yet computationally efficient, domain-expert models.

\begin{figure}[htbp]
    \centering
     \includegraphics[width=0.9\linewidth]{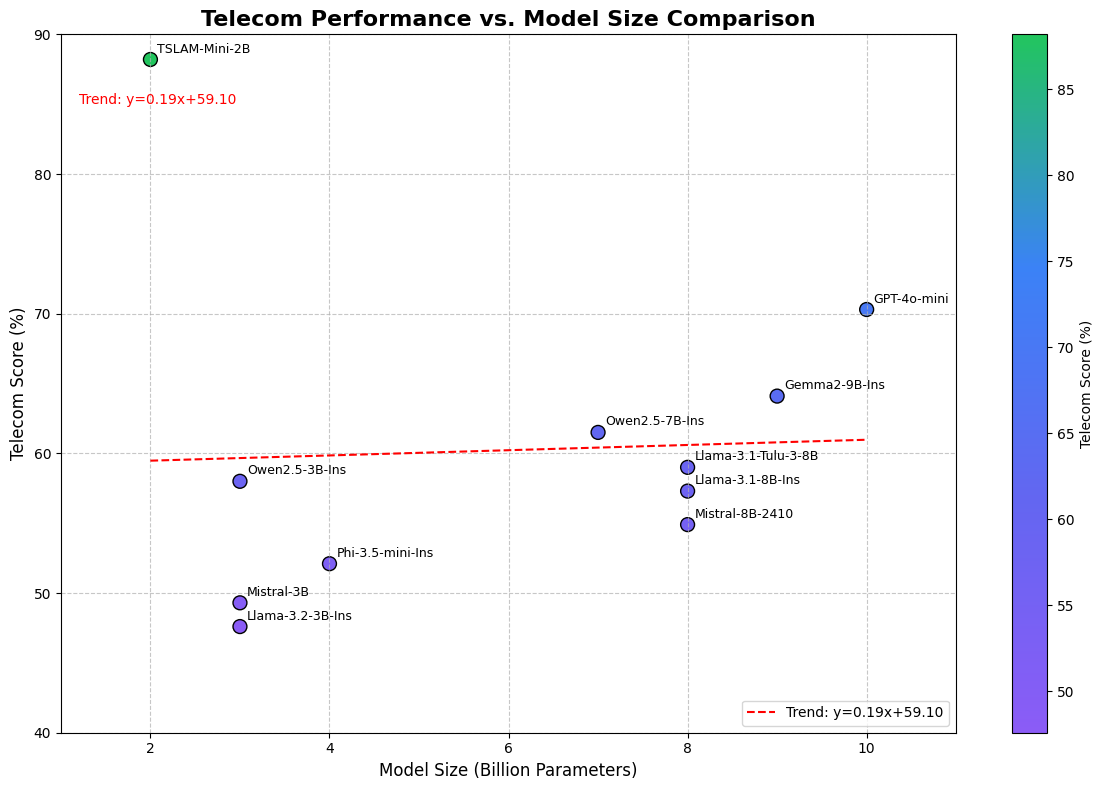}
    \caption{Comparative performance of TSLAM-Mini against selected larger general-purpose LLMs on telecom-specific evaluation benchmarks. TSLAM-Mini demonstrates strong performance on telecom domain tasks. }
    \label{fig:4.png}
\end{figure}

The model's performance parity on general tasks (when benchmarked against its base model, Phi-4 Mini Instruct 4B) combined with superior specialized performance indicates successful knowledge infusion without catastrophic forgetting of general capabilities. The results strongly advocate for the strategic development of smaller, domain-expert LLMs as a viable and efficient alternative to relying solely on monolithic, general-purpose models for specialized industrial applications.

\section{Future Work}
\label{sec:future_work}
The development of TSLAM-Mini establishes a robust foundation for specialized LLMs in telecommunications. Future research will concentrate on augmenting its reasoning faculties and optimizing fine-tuning methodologies.

\subsection{Enhancing Reasoning with Reinforcement Learning from Human Feedback (RLHF)}
To further elevate TSLAM-Mini's sophisticated reasoning capabilities, particularly in complex, multi-step diagnostic or optimization tasks, we propose the integration of Reinforcement Learning from Human Feedback (RLHF). This will involve:
\begin{itemize}
    \item \textbf{Reasoning-Driven Reward Modeling:} Developing reward models based on telecom expert feedback that specifically incentivize logical consistency, causal inference, and optimal decision-making in simulated network scenarios (e.g., fault localization pathways, network slice configuration trade-offs).
    \item \textbf{Context-Aware Prompt Refinement and Clarification:} Training TSLAM-Mini via RLHF to proactively seek clarification for ambiguous prompts or to infer implicit contextual parameters critical for telecom operations, thereby enhancing its utility and reliability in dynamic environments.
\end{itemize}
This initiative aims to align TSLAM-Mini’s outputs more closely with the nuanced judgment of human experts, particularly for tasks requiring intricate reasoning, such as optimizing 5G RAN parameters based on dynamic traffic patterns.

\subsection{Novel Quantization-Aware PEFT for Ameliorated Data Fidelity}
Building upon the QLoRA framework, we intend to investigate and develop advanced quantization-aware PEFT strategies. The objective is to further mitigate information loss typically associated with quantization, especially in ultra-low-precision regimes (e.g., sub-4-bit), while preserving or enhancing computational efficiency. This is paramount for deploying TSLAM-Mini on highly resource-constrained edge devices prevalent in modern telecom infrastructures. Key research thrusts include:
\begin{itemize}
    \item \textbf{Development of Advanced Quantization-Resilient PEFT Methodologies:} Exploring techniques such as quantization-aware training specifically for LoRA adapters, or hybrid quantization schemes that selectively apply different precision levels to various model components based on sensitivity analysis.
    \item \textbf{Empirical Validation on Edge-Representative Benchmarks:} Rigorously evaluating these novel PEFT strategies against telecom-specific benchmarks simulating edge deployment constraints, focusing on metrics like network configuration accuracy, real-time telemetry analysis fidelity, and fault diagnosis precision post-quantization.
\end{itemize}
These advancements are posited to yield a TSLAM-Mini variant with superior reasoning capabilities and heightened efficiency, poised to address the evolving complexities of intelligent network management.

\section{Conclusion}
\label{sec:conclusion}
TSLAM-Mini signifies a substantial contribution to the application of language models within the telecommunications domain, effectively addressing the inherent limitations of general-purpose LLMs through targeted domain-specific fine-tuning and computationally efficient training paradigms. The development of a comprehensive 100,000-sample dataset, synthesized via NetoAI’s DigiTwin platform and enriched with telecom-specific RFCs and SME expertise, endows TSLAM-Mini with a deep understanding of 20 critical telecommunications use-cases. The strategic employment of QLoRA facilitated high-performance fine-tuning of the compact Phi-4 Mini Instruct 4B architecture (3.8B parameters), enabling potential deployment on resource-constrained hardware while demonstrating performance competitive with significantly larger models on specialized tasks.

Our novel evaluation framework, incorporating the Qwen3-235B-A22B model as an automated adjudicator, provided a rigorous and objective assessment methodology. The empirical results affirm TSLAM-Mini's superior proficiency in telecom-specific applications, such as network optimization and technical query resolution, thereby validating the efficacy of specialized datasets and PEFT strategies. Prospective enhancements, including the integration of RLHF for advanced reasoning and the development of novel quantization-aware PEFT techniques, are poised to further refine TSLAM-Mini’s capabilities and operational efficiency. This research not only propels the application of LLMs in telecommunications but also establishes a blueprint for future investigations into domain-specific AI, showcasing the transformative potential of compact, expertly-tailored models in complex, real-time industrial sectors.



\end{document}